\documentclass[twocolumn,showpacs,preprintnumbers,amsmath,amssymb]{revtex4}

\usepackage{epsf}
\usepackage{graphicx}
\usepackage{amsfonts}
\usepackage{dcolumn}%
\usepackage{color,ulem}
\usepackage{bm}

\begin{document}

\preprint{}

\title{Long-range spin current driven by superconducting phase difference in a Josephson junction with double layer ferromagnets}

\author{S. Hikino$^{1}$}
\author{S. Yunoki$^{1,2,3}$}%
\affiliation{%
$^{1}$Computational Condensed Matter Physics Laboratory, RIKEN ASI, Wako, Saitama 351-0198, Japan \\
$^{2}$CREST, Japan Science and Technology Agency, Kawaguchi, Saitama 332-0012, Japan\\
$^{3}$Computational Materials Science Research Team, RIKEN AICS, Kobe, Hyogo 650-0047, Japan
} 

\date{\today}

\begin{abstract}
We theoretically study spin current through ferromagnet (F) in a Josephson junction composed of 
$s$-wave superconductors and two layers of ferromagnets. 
Using quasiclassical theory, we show that the long-range spin current can be driven by the superconducting phase difference without voltage drop. 
The origin of this spin current is due to spin-triplet Cooper pairs (STCs) formed by electrons of equal-spin, which are 
induced by proximity effect inside the F. 
We find that the spin current carried by the STCs exhibits long-range propagation in the F 
even where the Josephson charge current is practically zero. 
We also show that this spin current persists over a remarkably longer distance than the ordinary spin current carried by spin 
polarized conduction electrons in the F. 
Our results thus indicate the promising potential of Josephson junctions based on multilayer ferromagnets 
for spintronics applications with long-range propagating spin current. 
\end{abstract}

\pacs{74.45.+c, 72.25.Ba, 74.78.Na}
\maketitle

Spin current, a flow of electron spin angular momenta, can be generated in various ferromagnetic materials 
and plays a key role in spintronics. Spintronics devises have the advantage over the conventional electronics 
in data storage, non-evaporate memory, low power consumption, 
and high speed processing~\cite{zutic-rev, maekawa-book, handbook}.
These spintronics devises are controlled by spin current and thus the well defined spin current is of crucial importance. 
However, it is well known that the spin current carried by spin polarized electrons in ferromagnets can propagate only for a short 
distance. This 
is simply because the propagation distance of the spin current is determined by the spin diffusion length in ferromagnets, which 
is typically in a range of a few -- 10 nm~\cite{kimura,kimura-she}. 
Therefore, efficient generation of long-range propagating spin current is one of the primary issues 
in spintronics. 

It is notable that spintronics devises combined with superconductors have been rapidly developed for the last decade. 
The superconducting spintronics exhibits many fascinating phenomena which are not observed in the non-superconducting 
spintronics~\cite{yang,hubler}. The most fundamental element of the superconducting spintronics involves 
an $s$-wave superconductor/ferromagnet (S/F) hybrid structure. 
In a S/F junction, due to the proximity effect, spin singlet Cooper pairs (SSCs) penetrate into the F. 
Because of the exchange splitting of the electronic density of states for up and down electrons, the SSC in the F 
acquires the finite center-of-mass momentum, and the pair amplitude shows a damped oscillatory behavior with the thickness 
of the F~\cite{buzdin-sfs1}. 
For application purposes, the most interesting effect in the superconducting spintronics so far is the so called $\pi$-state in a 
S/F/S Josephson junction~\cite{buzdin-sfs1, buzdin82-91, ryazanov, kontos}. 
As opposed to the ordinary Josephson junction, i.e., S/normal metal/S, the $\pi$-state has the minimum Josephson 
coupling energy at the superconducting phase difference of $\pi$. 
It has been proposed that the $\pi$-state can be used for, e.g., quantum computing~\cite{yamashita-qbit}. 

One severe limitation in the superconducting spintronics devises based on the SSCs is, however, that 
the proximity effect in the S/F junction becomes negligibly small at a distance exceeding the magnetic 
coherence length $\xi_{\rm F}$, 
which is typically shorter than a few dozen nm~\cite{ryazanov,kontos,robinson}. 
Therefore, the penetration length of the SSC into the F is very short. This implies that any devise based on 
superconducting spintronics has to be smaller than nm size. 

By contrast, the spin triplet Cooper pairs (STCs), composed of electrons of equal spin with spin $|S|=1$, 
are superior because of the very long penetration depth into the F in the S/F junction. 
It is known that not only the SSCs but also the STCs can be 
induced in the F of the S/F junction, for example, when the F contains magnetic domain wall~\cite{ bergeret-prl86,bergeret-sfs1}, 
the F has spiral magnetic structure (as in Ho)~\cite{halasz-prb84}, the junction consists of multilayers of Fs~\cite{houzet,volkov-prb81}, 
or the interface is spin active~\cite{eschrig-sh,asano-prl}. 
Although the pair amplitude monotonically decreases with the thickness of the F, in these S/F junctions the STC can propagate into 
the F over a distance of the order of normal metal coherence length $\xi_{\rm N}$, 
which is typically about several hundred nm~\cite{deutscher, keizer,robinson-science, khaire, anwar-apl, anwar}. This is approximately two orders of magnitude 
larger than the propagation distance of the SSC ($\sim\xi_{\rm F}$). Thus, the proximity effect of the STC is called the 
long-range proximity effect. 

On the one hand, the charge transport of the STCs for these S/F junctions have been extensively studied. 
For instance, the Josephson current carried by the STCs has been predicted theoretically~\cite{bergeret-prl86,eschrig-sh} 
and confirmed experimentally~\cite{keizer, robinson-science, khaire, anwar-apl, anwar}. 
On the other hand, only limited studies have been reported thus far for the spin transport of the STCs in rather complex S/F junctions 
containing magnetic domain wall with spin active interfaces~\cite{alidoust, shomali}. 
Understanding the spin transport of the STCs remains highly desirable because transport properties in S/F multilayer systems 
strongly depend on the geometry of the junction as well as the property of F. 
The main purpose of our study is to propose a simple S/F junction, involving neither magnetic domain wall 
nor spin active interfaces, in which the proximity effect of the STC can induce the long-range propagating spin current. 


\begin{figure}[!t]
\begin{center}
\includegraphics[width=7cm]{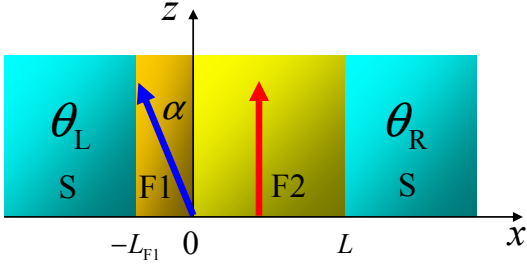}
\caption{(Color online) Schematic illustration of the S/F1/F2/S junction studied. 
Two arrows indicate the direction of magnetizations in F1 and F2 layers with thickness $L_{\rm F1}$ and 
$L$, respectively.  Here, the uniform magnetization is assumed in each F. 
$\alpha$ is the polar angle of the magnetization in the F1, while the magnetization direction in F2 is fixed along the $z$ direction. 
The phase difference between two 
$s$-wave superconductors is $\theta=\theta_{\rm R}-\theta_{\rm L}$. 
}
\label{sf1f2s}
\end{center}
\end{figure}

In this Letter, we study the spin current through the F of a Josephson junction composed of $s$-wave superconductors and 
two layers of ferromagnets with no spin active interfaces (Fig.~\ref{sf1f2s}). 
Based on the quasiclassical Green's function theory, we show that the long-range spin current can be driven by 
the superconducting phase difference ($\theta$) without voltage drop. 
The origin of this spin current is due to the STCs in the F induced by the proximity effect. 
We find that the spin current carried by the STCs can propagate over a much longer distance inside the F, as compared with 
the ordinary spin current carried by spin polarized conduction electrons, even where the 
Josephson current is practically zero. 
Our result therefore indicates that the spin and charge degrees of freedom can in practice be separated in the present S/F junction.

As depicted in Fig.$~$\ref{sf1f2s}, we consider the S/F1/F2/S 
junction made of two layers of ferromagnets (F1 and F2) attached to $s$-wave superconducting electrodes. 
We assume that each layer is good electric contact with the same mean free path 
and the same conductivity, and with no spin active interfaces. 
The spin current in the diffusive limit is evaluated by solving the linearized Usadel equation 
in each region~\cite{demler, bergeret-sfs1, buzdin-sfs1}, 
\begin{equation}
i\hbar D\partial ^{2}_{x} {\hat f} -i2\hbar | \omega_{n} | {\hat f} +2 {\hat \Delta} 
-{\rm sgn}(\omega_{n})  [{\vec h_{\rm ex}} \cdot {\hat \sigma},{\hat f}] 
=\hat 0
\label{usadel}, 
\end{equation}
where $D$ is the diffusion coefficient, $\omega_n$ is the fermion Matsubara frequency, 
${\rm sgn}(A)$ represents the sign of $A$, and 
$\hat \sigma $ is the Pauli matrix~{\cite{note5}. 
The anomalous part ${\hat f}$ of the $(2\times2)$ quasiclassical Green's function~\cite{eschrig-sh} is given by 
\begin{equation}
\hat f = \left( {\begin{array}{*{20}c}
   {f_{ \uparrow  \uparrow } } & {f_{ \uparrow  \downarrow } }  \\
   {f_{ \downarrow  \uparrow } } & {f_{ \downarrow  \downarrow } }  \\
\end{array}} \right) = \left( {\begin{array}{*{20}c}
   { - f_{tx}  + if_{ty} } & {f_s  + f_{tz} }  \\
   { - f_s  + f_{tz} } & {f_{tx}  + if_{ty} }  \\
\end{array}} \right)
\label{qgf}, 
\end{equation}
where $x$ dependence is implicitly assumed. The $s$-wave superconducting gap $\hat \Delta$ is 
finite only in the S and assumed to be constant, i.e., 
\begin{equation}
\hat \Delta  = \left\{ \begin{array}{l}
 \left( {\begin{array}{*{20}c}
   0 & { - \Delta }  \\
   \Delta  & 0  \\
\end{array}} \right),\,\,\,\,\,x < -L_{\rm F1},\,\,\,\,\,x > L  \\ 
 \hat 0,\,\,\,\,\,\,\,\,\,\,\,\,\,\,\,\,\,\,\,\,\,\,\,\,\,\,\,\,\,\,\,\,\,\,-L_{\rm F1} < x < L  \\ 
 \end{array} \right.
\label{gap}. 
\end{equation}
The exchange field ${\vec h}_{\rm ex}$ due to the 
ferromagnetic magnetizations in the Fs 
is described by 
\begin{equation}
\vec h_{\rm ex}  = \left\{ \begin{array}{l}
 0, \,\,\,\,\,\,\,\,\,\,\,\,\,\,\,\,\,\,\,\,\,\,\,\,\,\,\,\,\,\,\,\,\,\,\,\,\,\,\,\,\,\,\,\,\,\,\,\,\,\,\,\,\,\,\,\,\,\,\,\,\,\,x < -L_{\rm F1} \\ 
 h_{\rm ex1} \left( {-\sin \alpha \vec e_x  + \cos \alpha \vec e_z } \right), -L_{{\rm{F1}}} < x < 0  \\ 
 h_{\rm ex2}\vec e_z,\,\,\,\,\,\,\,\,\,\,\,\,\,\,\,\,\,\,\,\,\,\,\,\,\,\,\,\,\,\,\,\,\,\,\,\,\,\,\,\,\,\,\,\,\,\,\,\,\,0 < x < L  \\ 
 0, \,\,\,\,\,\,\,\,\,\,\,\,\,\,\,\,\,\,\,\,\,\,\,\,\,\,\,\,\,\,\,\,\,\,\,\,\,\,\,\,\,\,\,\,\,\,\,\,\,\,\,\,\,\,\,\,\,\,\,\,\,\,x > L  \\ 
 \end{array} \right.
\label{hex},  
\end{equation}
where $\alpha$ is the polar angle of the magnetization in the F1 and $\vec e_{x(z)}$ is the unit vector along the $x(z)$ 
direction (see Fig.~\ref{sf1f2s}).  
Within the quasiclassical theory and the linearized approximation, 
the spin current polarized in the $x$ spin quantization axis 
flowing along the $x$ direction perpendicular to the junction 
is given by 
\begin{eqnarray}
	j_{\rm S}^{\bot } (x) &=& \frac{ieDN(0)}{2\beta}
	\sum_{i\omega_{n}}
	{\rm tr} 
	\left[
	\hat \sigma^{x} (\hat f \partial_{x} \hat f^{\dagger } -\hat f^{\dagger } \partial_{x} \hat f ) 
	\right], \nonumber \\
	&=&
	\frac{2eDN(0)}{\beta}
	\sum_{i\omega_{n}} {\rm Im}
	\left(
	f_{tx}\partial_{x}f_{s}^{\dagger}-f_{s}\partial_{x}f_{tx}^{\dagger}
	\right. \nonumber \\
	&&\left.
	\quad\quad\quad\quad\quad\quad+ if_{tz}\partial_{x}f_{ty}^{\dagger}-if_{ty}\partial_{x}f_{tz}^{\dagger}
	\right)
\label{spin-current1}, 
\end{eqnarray}
where $N(0)$ is the density of states (DOS) per unit volume and per electron spin at the Fermi energy~\cite{note4},  
$\beta=1/k_{\rm B}T$, and $T$ is temperature. In the following, we calculate the spin current flowing through the F2 
using the above equation.

The Usadel equation given in Eq.~(\ref{usadel}) can be solved by imposing appropriate boundary 
conditions~\cite{demler, houzet}. 
Here, we assume that the superconducting electrodes are much larger than the ferromagnetic 
layers. Then, obviously, 
the anomalous Green's function in the right (left) S for $x\rightarrow  +\infty$ $(-\infty)$ is given by 
$f_{s}=|\Delta|e^{i \theta_{\rm R(L)}}/\hbar|\omega_{n}|$, 
i.e., the solution of the Usadel equation in a bulk S. 
Solving Eq.$~$(\ref{usadel}) with these boundary conditions, the anomalous Green's function ${\hat f}^{\rm F2}$ in the F2 is given by 
\begin{equation}
	f_{s}^{\rm F2} =
	-i\frac{\Delta_{\rm R} \kappa_{\rm F2} \Phi_{\omega_{n}}^{\rm s} }{\hbar |\omega_{n}| k }
	\Pi(x) 
	-i\frac{\Delta_{\rm L} \kappa_{\rm F2} \Phi_{\omega_{n}}^{\rm s} }{\hbar |\omega_{n}| k }
	\Xi(x)
\label{fs}, 
\end{equation}
and 
\begin{eqnarray}
	f_{tx}^{{\rm{F2}}}  &=& 
	\frac{\Delta_{\rm R}}{\hbar |\omega_{n}| }
	\frac{{\rm sgn}(\omega_{n}) h_{\rm ex1}^{x} \kappa_{\rm F1}^{2}\kappa_{\rm F2}L_{\rm F1} \Phi_{\omega_{n}}^{\rm s}}
	{\hbar D k^{2} (\kappa_{\rm F1}-k)(\kappa_{\rm F1}+k)}
	e^{-kx} \nonumber \\
	&+&
	\frac{\Delta_{\rm L}}{\hbar |\omega_{n}| }
	\frac{{\rm sgn}(\omega_{n}) h_{\rm ex1}^{x} \kappa_{\rm F1}^{2}\kappa_{\rm F2}L_{\rm F1} \Phi_{\omega_{n}}^{\rm s}}
	{\hbar D k^{2} (\kappa_{\rm F1}-k)(\kappa_{\rm F1}+k)}
	\Pi(L)
	e^{-kx}
\label{ftx}, 
\end{eqnarray}
where $\Delta_{\rm R(L)}=|\Delta|e^{i\theta_{\rm R(L)}}$, 
\begin{equation}
	{\Phi_{\omega_{n}}^{\rm s}} =
	\left\{
	2\frac{\kappa_{\rm F2 }}{k} \cosh(\kappa_{\rm F2} L) 
	+
	\left[
	1+
	\left(
	\frac{\kappa_{\rm F2}}{k}
	\right)^{2}
	\right]
	\sinh(\kappa_{\rm F2} L)
	\right\}^{-1}
\label{phis}	\nonumber ,
\end{equation}
and 
\begin{equation}
\left.
\begin{array}{r}
\Pi(x)\\
\Xi(x+L)
\end{array}
\right\} 
= 	\cosh\left(\kappa_{\rm F2} x \right) 
	\pm \frac{k}{\kappa_{\rm F2}} \sinh\left(\kappa_{\rm F2} x \right) 
\nonumber
\end{equation}
with $\kappa _{{\rm{F1}}\left( {{\rm{F2}}} \right)}  = \sqrt{\frac { {2\hbar \left| {\omega _n } \right| - i2{\mathop{\rm sgn}} \left( {\omega _n } \right)
h_{\rm ex1} \cos \alpha \left( {h_{\rm ex2} } \right)}}{ \hbar D } }$, $k=\sqrt{ 2 |\omega_{n}|/D }$, and 
$h_{\rm ex1}^{x}=-h_{\rm ex1}\sin\alpha$. 
Notice that $k^{-1}$ with $n=0$ corresponds to the normal metal coherence length $\xi_{\rm N}=\sqrt{\frac{\hbar D}{2 \pi k_{\rm B} T}}$. 
It should be noted that $f_{ty}=0$ since the exchange field in the F1 does not have the $y$ component. 
Eqs.$~$(\ref{fs}) and (\ref{ftx}) are obtained assuming $\kappa _{{\rm{F1}}\left( {{\rm{F2}}} \right)}L_{\rm F1}$, 
$k L_{\rm F1}$ $\ll 1$~{\cite{note2}. 
Inserting Eqs.$~$(\ref{fs}) and (\ref{ftx}) into Eq.$~$(\ref{spin-current1}) and 
$L$ into $x$, we obtain the spin current, 
$j_{\rm S}^{\bot }(L,\theta)=j_{\rm SC}^{\bot}(L)\sin\theta$, flowing through the F2. 
Here, $j_{\rm SC}^{\bot}(L)=j_{\rm SC1}^{\bot }(L)+j_{\rm SC2}^{\bot }(L)$ is the $\theta$ independent part, and 
$j_{\rm SC1}^{\bot }(L)$ and $j_{\rm SC2}^{\bot }(L)$ are given by 
\begin{eqnarray}
	j_{\rm SC1(2)}^{\bot }(L) &=& -\frac{2e\pi N(0) h_{ex1}^{x}}{\hbar \beta}
	\sum_{i \omega_{n}} 
	{\rm Im}
	\left[
	\frac{|\Delta \kappa_{\rm F2}|}{\hbar |\omega_{n}|}^{2}
	\frac{ {\rm sgn}(\omega_{n})}{k}
	\right.  
	\nonumber \\
	&\times &
	\left.
	|\Phi_{\omega_{n}}^{\rm s}|^{2}
	\Phi_{1(2)}(L,\omega_{n} )
	e^{-kL}
	\right]
\label{jsc12}.
\end{eqnarray}
The $L$ dependent functions $\Phi_{1(2)}(L,\omega_{n} )$ in Eq.~(\ref{jsc12}) are defined by 
\begin{widetext}
\begin{eqnarray}
	\Phi_{1}(L,\omega_{n}) &=&
	\frac{\kappa_{\rm F1}}{k} \frac{\kappa_{\rm F1} L_{\rm F1} }{{\left( {\kappa _{{\rm{F1}}}  - k} \right)\left( {\kappa _{{\rm{F1}}} 
	+ k} \right)}}\left[ {1 + \frac{1}{2}\left( {1 + \frac{{\kappa _{{\rm{F2}}}^{\rm
	{*}} }}{{\kappa _{{\rm{F2}}} }}} \right)\cosh \left( {2k_1^{{\rm{F2}}} L} \right
	) + \frac{1}{2}\left( {\frac{{\kappa _{{\rm{F2}}}^{\rm{*}} }}{k} + \frac{k}{{
	\kappa _{{\rm{F2}}} }}} \right)\sinh \left( {2k_1^{{\rm{F2}}} L} \right)} \right] \nonumber \\
	&-&
	\frac{\kappa_{\rm F1}^{*}}{k} \frac{\kappa_{\rm F1}^{*} L_{\rm F1} }{{\left( {\kappa _{{\rm{F1}}}^{\rm{*}}  - k} \right)\left( 
	{\kappa _{{\rm{F1}}}^{\rm{*}}  + k} \right)}}\left\{ {1 - \frac{1}{2}\left[ {1 + 
	\left( {\frac{k}{{\left| {\kappa _{{\rm{F2}}} } \right|}}} \right)^2 } \right]\cosh \left( {
	2k_1^{{\rm{F2}}} L} \right) - {\mathop{\rm Re}\nolimits} \left( {\frac{k}{
	{\kappa _{{\rm{F2}}} }}} \right)\sinh \left( {2k_1^{{\rm{F2}}} L} \right)} \right\} 
	\label{phi12_1}
\end{eqnarray}
and
\begin{eqnarray}
	\Phi_{2}(L,\omega_{n}) &=&
	\frac{\kappa_{\rm F1} }{k}\frac{\kappa_{\rm F1} L_{\rm F1}}{{2\left( {\kappa _{{\rm{F1}}}  - k} \right)\left( {\kappa _{{\rm{F1}}} 
	 + k} \right)}}\left[ {\left( {1 - \frac{{\kappa _{{\rm{F2}}}^{\rm{*}} }}
	 {{\kappa _{{\rm{F2}}} }}} \right)\cos \left( {2k_2^{{\rm{F2}}} L} \right) +
	 i{\mathop{\rm sgn}} \left( {\omega _n } \right)\left( {\frac{{\kappa _{{\rm{F2}}}^{\rm{*}}
	 }}{k} - \frac{k}{{\kappa _{{\rm{F2}}} }}} \right)\sin \left( {2k_2^{{\rm{F2}}} L} \right)} \right] \nonumber \\
	 &+&
	\frac{\kappa_{\rm F1}^{*}}{k} \frac{\kappa_{\rm F1}^{*} L_{\rm F1} }{{2\left( {\kappa _{{\rm{F1}}}^{\rm{*}}  - k} \right)\left( 	 
	{\kappa _{{\rm{F1}}}^{\rm{*}}  + k} \right)}}\left\{ {\left[ {1 - \left( {\frac{k}{
	{\left| {\kappa _{{\rm{F2}}} } \right|}}} \right)^2 } \right]\cos \left( {2k_2^{{\rm{F2}}} L} \right) 
	+ 2{\mathop{\rm sgn}} \left( {\omega _n } \right)
	{\mathop{\rm Im}\nolimits} \left( {\frac{k}{{\kappa _{{\rm{F2}}} }}} \right)
	\sin \left( {2k_2^{{\rm{F2}}} L} \right)} \right\}
\label{phi12_2}
\end{eqnarray}
\end{widetext}
with $k_{1(2)}^{\rm F2} = 
	\sqrt {\frac{{\sqrt {\left( {\hbar \left| {\omega _n } \right|} \right)^2 
	+ \left( { h_{\rm ex2} } \right)^2 }  +(-) \hbar \left| {\omega _n } \right|}}{{\hbar D}}}$. 
It should be readily noticed in Eqs.~(\ref{jsc12})--(\ref{phi12_2}) that $j_{\rm SC1}^{\bot}  (L)$ decreases monotonically 
with $L$, whereas $j_{\rm SC2}^{\bot}  (L)$ shows damped oscillatory behavior. 
Although Eq.~(\ref{jsc12}) is obtained by 
products of $f_{tx}$ and $f_s$ [see Eq.~(\ref{spin-current1})], this observation indicates that 
the main contribution to $j_{\rm SC1}^{\bot}  (L)$ [$j_{\rm SC2}^{\bot}  (L)$] is the STC (SSC).  
This is also supported by the fact that $j_{\rm SC1}^{\bot}  (L)$ [$j_{\rm SC2}^{\bot}  (L)$] is insensitive (sensitive) to $h_{{\rm ex}2}$. 
	
The Josephson current flowing through the F2 is similarly calculated. 
Based on the quasiclassical Green's function theory and the linearized approximation, the Josephson current is given by 
\begin{equation}
	j_{\rm J} (x)= 
	\frac{i \pi e D N(0)}{2 \beta}
	\sum_{i \omega_{n}}
	{\rm tr}
	\left(
	\hat f
	\partial _{x}
	\hat f ^\dagger
	-
	\hat f ^\dagger 
	\partial_{x}
	\hat f
	\right)
\label{j}.
\end{equation}
Substituting into Eq.$~$(\ref{j}) the solutions of Eq.$~$(\ref{usadel}) with the appropriate boundary conditions, 
we obtain the Josephson current $j_{\rm J}(L,\theta) = j_{\rm c}(L) \sin \theta$ with 
%
\begin{eqnarray}
	j_{\rm c}(L) &=& 
	\frac{2 \pi e D N(0)}{\beta}
	\sum_{i \omega_{n}}
	\left(
	\frac{ |\Delta \kappa_{\rm F2}| }{ \hbar |\omega_{n}| }
	\right)^{2}
	\frac{1}{k} \nonumber \\
	&\times&
	{\rm Im}
	\left[
	i \cosh(\kappa_{\rm F2}L) 
		\left[
		1+\frac{k}{\kappa_{\rm F2}} \tanh(\kappa_{\rm F2}L)
		\right] 
	\right. \nonumber \\
	&+& 
	\left.
	i \cosh(\kappa_{\rm F2}^{*}L)
		\left[
		1+\frac{ \kappa_{\rm F2}^{*} }{k} \tanh(\kappa_{\rm F2}^{*}L)
		\right]
	\right]
	|\Phi_{\omega_{n}}^{\rm s}|^{2}
\label{jc}. \nonumber \\
\end{eqnarray}
Here, $\kappa _{{\rm{F1}}\left( {{\rm{F2}}} \right)}L_{\rm F1}$, $k L_{\rm F1}$ $\ll $1 is assumed as in the derivation of the spin current. 
It should be noted that in the present system the Josephson current is carried solely by the SSCs, but not by the STCs. 
This is because the spin triplet contribution ${\rm Im}(f_{tx} \partial_{x} f_{tx}^{\dagger})$ is 
exactly zero in Eq.$~$(\ref{ftx}). 
In order for the STCs to contribute the non-zero Josephson current, 
a Josephson junction with trilayer ferromagnets, for example, is necessary, as pointed out in Ref.~\cite{houzet}. 

\begin{figure}[!t]
\begin{center}
\includegraphics[width=7.3cm]{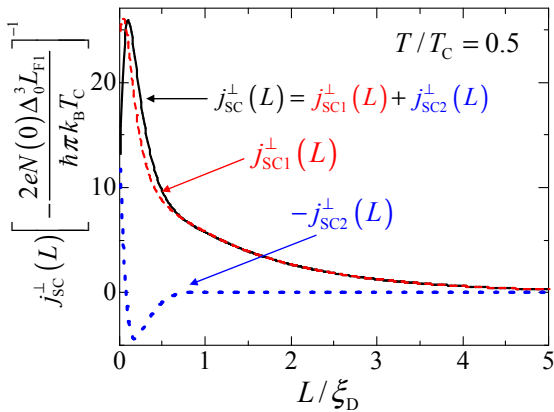}
\caption{(Color online) Spin current $j_{\rm SC}^{\bot }(L)$ flowing through the F2 as a function of thickness $L$ of the F2. 
Here, $\xi_{\rm D}$ is the normal metal coherence length $\xi_{\rm N}$ at $T=T_{\rm C}$ 
and $N(0)$ is the DOS for the F2. 
The parameters used are $h_{\rm ex1}/\Delta_{0}=30$, 
$h_{\rm ex2}/\Delta_{0}=20$, $\alpha=-\pi/3$, and $T/T_{\rm C}=0.5$.  For comparison, $j_{\rm SC1}^{\bot }(L)$ and 
$j_{\rm SC2}^{\bot }(L)$ [Eq.~(\ref{jsc12})] are also plotted, separately. 
}
\label{js-d}
\end{center}
\end{figure}

Let us now evaluate numerically the spin current in the F2. For this purpose, 
the temperature dependence on $\Delta$ is assumed to be $\Delta=\Delta_{0} \tanh(1.74 \sqrt{T_{\rm C}/T - 1})$, where  
$\Delta_{0}$ and $T_{\rm C}$ are the superconducting gap at zero temperature and 
the superconducting transition temperature, respectively~\cite{belzig-gap}. 
Figure~\ref{js-d} shows a typical result of the spin current as a function of thickness $L$ of the F2 normalized 
by the normal metal coherent length $\xi_{\rm N}$ at $T=T_{\rm C}$ (denoted by $\xi_{\rm D}$). 
We find in Fig.$~$\ref{js-d} that the spin current, $j_{\rm SC}^{\bot }(L)=j_{\rm SC1}^{\bot }(L)+j_{\rm SC2}^{\bot }(L)$, 
decreases monotonically with $L$ for $L>\xi_D$ but can 
propagate over a much longer distance than the Josephson current for the same system (as will be shown below). 
It is also clear in Fig.$~$\ref{js-d} that the long-range propagating spin current $j_{\rm SC}^{\bot }(L)$ 
is originated mostly from $j_{\rm SC1}^{\bot }(L)$, which is carried by the STCs.  
Instead, the other component $j_{\rm SC2}^{\bot }(L)$ of the spin current shows strongly damped oscillatory behavior with $L$. 
This is easily understood because $j_{\rm SC2}^{\bot }(L)$ is carried by the SSCs, which are strongly destroyed in the F by the exchange field. 
Therefore, the origin of the long-range propagating spin current found here is attributed to the long-range proximity 
effect of the STC~\cite{note1}. 

Next, we compare the $L$ dependence on the spin and Josephson currents for different exchange field $h_{\rm ex2}$ in Fig.~\ref{js-jc-d}. 
These results in Fig.~\ref{js-jc-d} clearly reveal the long-range propagation characteristics of the spin current $j_{\rm SC}^{\bot }(L)$ 
as compared with 
the Josephson current $j_{\rm c}(L)$. 
Moreover, we find that the spin current $j_{\rm SC}^{\bot }(L)$ is insensitive to the exchange field $h_{\rm ex2}$. 
This is because the most of the spin current is carried by the STCs (see Fig.~\ref{js-d}), which are not destroyed by 
the exchange field. 
Instead, as seen in Fig.~\ref{js-jc-d}, the Josephson current $j_{\rm c}(L)$ strongly decreases with increasing 
$h_{\rm ex2}$. This strong decrease of $j_{\rm c}(L)$ with $h_{\rm ex2}$ is due to the pair breaking effect of the SSC by the exchange field. 
It should be also emphasized in Fig.~\ref{js-jc-d} that the spin current can propagate over a long distance even where the Josephson 
current is negligibly small for $L/\xi_{\rm D}\gg 1$. 
This result thus indicates that the spin and charge degrees of freedom can in practice be separated in this system, which is 
in sharp contrast to the previous studies in Refs.~\cite{alidoust} and \cite{shomali}. 

\begin{figure}[thbp]
\begin{center}
\includegraphics[width=7cm]{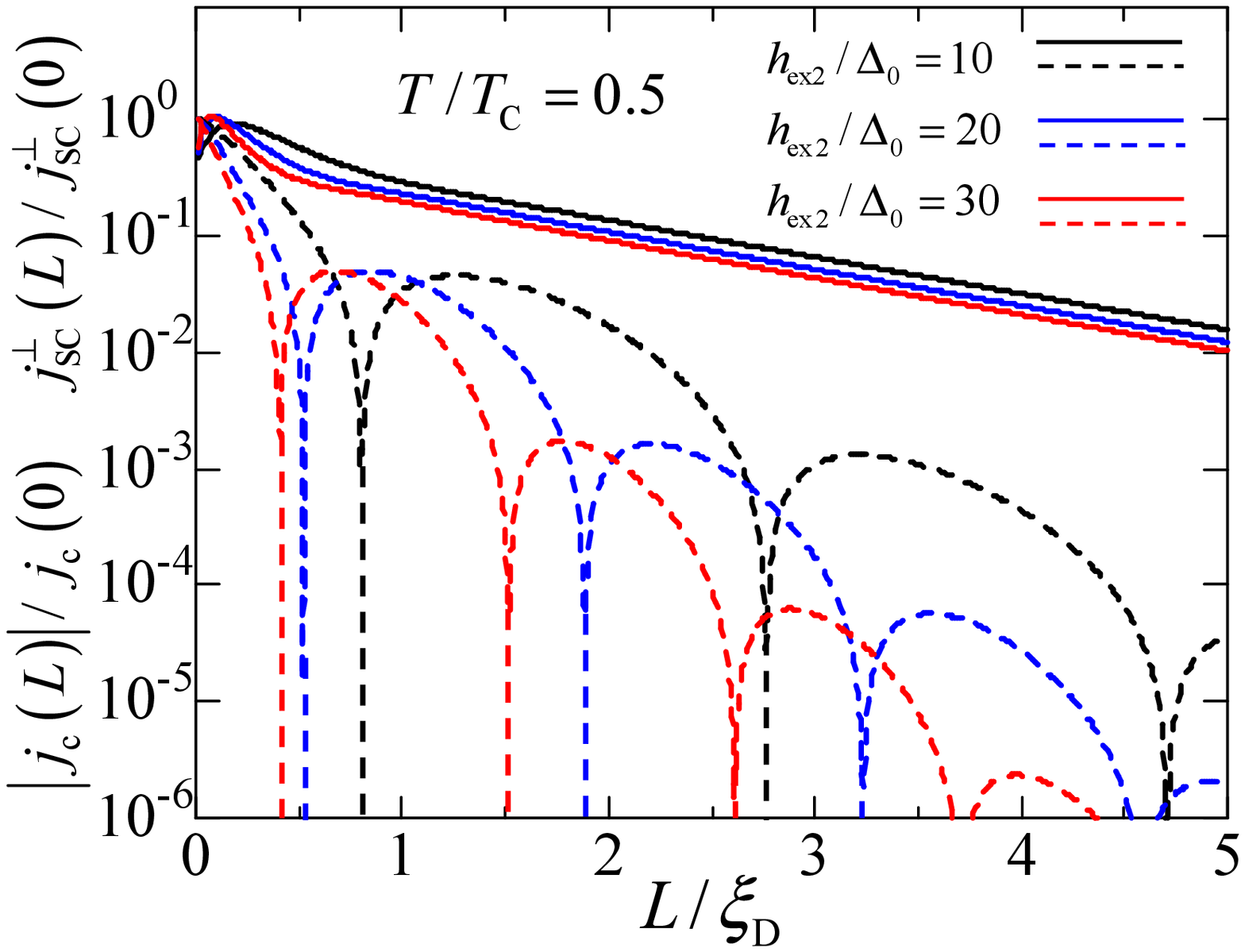}
\caption{(Color online) Spin current $j_{\rm SC}^{\bot }(L)$ (sold lines) and Josephson 
current $j_{\rm c}(L)$ (dashed lines) flowing through the F2 as a function of thickness $L$ of the F2 for 
different $h_{\rm ex2}/\Delta_{0}$ indicated in the figure. 
Here, $\xi_{\rm D}$ is the normal metal coherence length $\xi_{\rm N}$ at $T=T_{\rm C}$. 
The other parameters used are $h_{\rm ex1}/\Delta_{0}=30$, $\alpha=-\pi/3$, and $T/T_{\rm C}=0.5$.  
Note that the oscillatory behavior of the Josephson current is due to the sign change. 
}
\label{js-jc-d}
\end{center}
\end{figure}

Let us now approximately estimate the propagating distance and the amplitude of the spin current inside the F2. 
In the case of dirty metal, $\xi_{\rm D}$ is in a range of several dozen - several
hundred nm~\cite{deutscher, keizer,robinson-science,khaire,anwar-apl,anwar}. 
As shown in Fig.~\ref{js-jc-d}, the spin current can flow inside the F2 up to this length scale. 
This already implies that this spin current, which is carried mostly by the STCs, can flow much longer than the ordinary spin current 
carried by spin polarized conduction electrons in the F 
because the latter disappears at a distance of the spin diffusion length which is about a few nm~\cite{kimura,kimura-she}. 
We thus predict that the long-range spin current carried by the STCs can propagate about 10--100 times longer than the ordinary spin current. 
The amplitude $A_{\rm SC}$ of the long-range spin current is estimated to be of order
$2eN(0) \Delta_{0}^{3}L_{\rm F1}/(\hbar \pi k_{\rm B} T_{\rm C})$ (see Fig.~\ref{js-d}). 
When we use a typical set of parameters~\cite{note3}, 
the amplitude $A_{\rm SC}$ is approximately $ 10^{9}$ A/$\rm{m}^2$, which should be large enough to be observed experimentally.  

Finally, we shall comment on how to experimentally detect the spin current in our system. 
Among several currently available experimental methods, the spin Hall effect (SHE) is the most likely candidate to observe 
the spin current in the F. 
Indeed, recently, the SHE in a nonmagnetic Josephson junction has been theoretically predicted~\cite{Mal'shukov}. Similarly, 
the SHE is expected in our magnetic Josephson junction. 
Therefore, we expect that SHE devises~\cite{jungwirth} using the conventional experimental method~\cite{she} can detect and 
take out the spin current because Cooper pairs are generally scattered by spin dependent interactions~\cite{Mal'shukov}.
The detailed calculations of the SHE will be reported in the future. 

In summary, we have theoretically studied the spin current through the F in the Josephson junction 
composed of $s$-wave superconductors and two layers of ferromagnets. 
Based on the quasiclassical Green's function theory in the diffusive transport region, 
we have found that the long-range spin current can be driven by the superconducting phase difference without voltage drop. 
The origin of this spin current is due to the STCs induced by the proximity effect in the ferromagnet. 
We have shown that this spin current can propagate in the ferromagnet over a distance about 10--100 times longer than the ordinary 
spin current carried by spin polarized electrons, even where the Josephson current is practically zero. 
Our results clearly demonstrate that Josephson junctions based on multilayer ferromagnets can provide the new spin dependent 
transport and suggest the promising potential of these junctions for spintronics applications. 
 
S.~H. was supported in part by Japan Society for the Promotion of Science (JSPS).


\begin{thebibliography}{}
\bibitem{note2}
Note that with the assumption used here, i.e, $\sigma_{\rm F2}\approx \sigma_{\rm S}$ 
and $D_{\rm F2}\approx D_{\rm S}$, the coefficient $B$ of the term proportional to $e^{kx}$ in 
$f_{tx}^{\rm F2}$ [Eq.~(\ref{ftx})] is zero since  
$B\propto\left(1-\frac{\sigma_{\rm F2}}{\sigma_{\rm S}} \sqrt{\frac{D_{\rm S}}{D_{\rm F2}}}\right)$, 
where $\sigma_{\rm S(F2)}$ and $D_{\rm S(F2)}$ are the conductivity and the diffusion coefficient of S (F2), respectively.
\bibitem{note1}
Note that 
the spin current in the S/F1/F2/S junction as a whole is not conserved because the STC is destroyed inside 
the S when the thickness of the S is much longer than the superconducting coherent length 
$\xi_S$~\cite{bergeret-sfs1,eschrig-sh}. The properties of the STCs inside the S in a S/F multilayer 
system have been discussed recently by N. G. Pugach and A. I. Buzdin [Appl. Phys. Lett. {\bf 101}, 242602 (2012)].
\bibitem{note3}
A set of parameters used is $\Delta_{0}=1$ meV, $L_{\rm F1} = 5$ nm, $T_{\rm C}=9$ K, and 
Fermi energy $E_{\rm F} = 1$ eV. 
The free electron density of states is also assumed, i.e., 
$N(0)=\frac{1}{4 \pi^2} \left( \frac{2m}{ \hbar^{2}} \right)^{3/2} E_{\rm F}^{1/2}$.
\end{thebibliography}
\end{document}